\documentclass[aps,twocolumn,groupedaddress,showpacs,amsmath,amssymb]{revtex4}

\usepackage{graphicx}
\usepackage{dcolumn}
\usepackage{bm}

\begin{document}

\title{Polymorphism and superconductivity in the V-Nb-Mo-Al-Ga high-entropy alloys}

\date{\today}
\author{Jifeng Wu$^{1,2}$}
\author{Bin Liu$^{1,2}$}
\author{Yanwei Cui$^{2,3}$}
\author{Qinqing Zhu$^{1,2}$}
\author{Guorui Xiao$^{2,3}$}
\author{Hangdong Wang$^{4}$}
\author{Siqi Wu$^{3}$}
\author{Guang-han Cao$^{3}$}
\author{Zhi Ren$^{2}$}
\email{zhi.ren@wias.org.cn}

\affiliation{$^{1}$Department of Physics, Fudan University, Shanghai 200433, P. R. China}
\affiliation{$^{2}$School of Science, Westlake Institute for Advanced Study, Westlake University, Hangzhou 310064, P. R. China}
\affiliation{$^{3}$Department of Physics, Zhejiang University, Hangzhou 310027, P. R. China}
\affiliation{$^{4}$Department of Physics, Hangzhou Normal University, Hangzhou 310036, P. R. China}

\begin{abstract}
High-entropy alloys (HEAs) are at the focus of current research for their diverse properties, including superconductivity and structural polymorphism. However, the polymorphic transition has been observed only in nonsuperconducting HEAs and mostly under high pressure. Here we report the discovery of superconductivity and temperature-driven polymorphism in the (V$_{0.5}$Nb$_{0.5}$)$_{3-x}$Mo$_{x}$Al$_{0.5}$Ga$_{0.5}$ (0.2 $\leq$ $x$ $\leq$ 1.4) HEAs. It is found that the as-cast HEA is of a single body-centered cubic (bcc) structure for $x$ = 0.2 and a mixture of the bcc and A15 structures for higher $x$ values. Upon annealing, the bcc structure undergoes a polymorphic transformation to the A15 one and all HEAs exhibits bulk superconductivity. For $x$ = 0.2, whereas the bcc polymorph is not superconducting down to 1.8 K, the A15 polymorph has a superconducting transition temperature $T_{\rm c}$ of 10.2 K and and estimated zero-temperature upper critical field $B_{\rm c2}$(0) of 20.1 T, both of which are the highest among HEA superconductors. With increasing Mo content $x$, both $T_{\rm c}$ and $B_{\rm c2}$(0) of the A15-type HEAs decrease, yet the large ratio of $B_{\rm c2}$(0)/$T_{\rm c}$ signifies a disorder-induced enhancement of the upper critical field over a wide $x$ range. The decrease in $T_{\rm c}$ is attributed to the decrease in both the electronic specific-heat coefficient and electron-phonon coupling strength. Furthermore, the valence electron count dependence of $T_{\rm c}$, which is different from both the binary A15 and other structurally different HEA superconductors, suggests that $T_{\rm c}$ may be increased further by reducing the number of valence electrons. Our results not only uncover HEA superconductors of a new structural type, but also provide the first example of polymorphism dependent superconductivity in HEAs.
\end{abstract}

\maketitle
\maketitle
\section{Introduction}
In materials science, polymorphism refers to the property of a solid material to crystallize in at least two distinct structures. For superconductors, the presence of polymorphs is of considerable interest since it offers a unique opportunity to study how the spatial atomic arrangement affects the superconducting properties without changing the chemical composition, which may provide useful clues to the superconducting mechanism \cite{1C60}. So far, the concurrence of structural polymorphism and superconductivity at ambient pressure has been observed in pure elements \cite{39element}, binary alloys \cite{2binaryalloy1,3binaryalloy2}, ternary rare/alkali-earth intermetallics \cite{4rareearth1,5rareearth2}, perovskite oxides \cite{6oxide}, doped fullerides \cite{1C60}, transition metal dichalcogenides \cite{7chalcogenide}, and organic charge-transfer salts \cite{8organic1,9organic2}, all of which are ordered materials.

Recently, HEAs have received much attention due to their fascinating mechanical, thermal and physical properties \cite{10HEAreview1,11HEAreview2,12HEAreview3,13HEAreview4,14HEAreview5}. These alloys are single solid-solution phases made up of five or more metal elements, whose concentrations are restricted to between 5\% and 35\% atomic percent. Due to the extremely high chemical disorder, HEAs can be viewed as a metallic glass on an ordered lattice, and have been found mostly in high symmetry crystal structures, such as bcc, face-centered cubic (fcc), and hexagonal close packing (hcp). Among the large number of HEAs studied, a few members have been reported to show bulk type-II superconductivity \cite{15HEASCreivew}. These HEA superconductors are based on $d$ transition metal elements, and can be categorized into four different structural types: bcc type \cite{16bccHEASC}, $\alpha$-Mn type \cite{17alphaMnHEASC}, CsCl type \cite{18CsClHEASC}, and hcp type \cite{19hexagonalHEASC}. In particular, CsCl-type (ScZrNb)$_{0.65}$(RhPd)$_{0.35}$ HEA has the highest $T_{\rm c}$ of $\sim$9.3 K \cite{18CsClHEASC}, and the highest $B_{\rm c2}(0)$ of $\sim$11.7 T is achieved in the (TaNb)$_{0.5}$(ZrHfTi)$_{0.5}$ HEA with a simple bcc structure \cite{20HEABc2}. Moreover, the bcc-type (TaNb)$_{0.67}$(HfZrTi)$_{0.33}$ HEA shows robust zero-resistance superconductivity without structural transition at pressures up to 190 GPa, demonstrating the potential of HEAs for application under extreme conditions \cite{21HEAunderp}. However, no polymorphs of these HEA superconductors have been reported so far. As a matter of fact, the polymorphic transitions are observed only in nonsuperconducting HEAs and mostly at very high pressures, though the different polymorphs can still exist after the pressure is relieved \cite{22polymorphic1,23polymophic2,24polymorphic3}.

It is known that the formation energies of the bcc and A15 phases are very similar in some $A$$_{3}$$B$-type compounds, where $A$ is a group VB or VIB transition metal element such as V, Nb, Mo and $B$ is usually a IIIA main group element such as Al, Ga \cite{2binaryalloy1,25Nb3Al-1,26Nb3Al-2}. Consequently, a bcc-to-A15 polymorphic phase transition can be induced by thermal annealing. It is therefore of interest to study the HEAs based on the V-Nb-Mo-Al-Ga system, which, however, have not been explored to date. In this paper, we present a systematic study on the structural and physical properties of the (V$_{0.5}$Nb$_{0.5}$)$_{3-x}$Mo$_{x}$Al$_{0.5}$Ga$_{0.5}$ HEAs for 0.2 $\leq$ $x$ $\leq$ 1.4. It is shown that, the arc-melted (as-cast) HEA with $x$ = 0.2 possesses a disordered bcc structure, which transforms to the A15 structure after annealing at 1600 $^{\circ}$C. Whereas the bcc polymorph remains normal down to 1.8 K, the A15 one turns out to be a bulk superconductor with a $T_{\rm c}$ of 10.2 K and an orbitally-limited $B_{\rm c2}$(0) of 20.1 T. For $x$ $\geq$ 0.4, the as-cast HEAs contain a mixture of bcc and A15 polymorphs, and a similar bcc-to-A15 polymorphic transition is observed upon annealing. With increasing Mo content $x$, both $T_{\rm c}$ and $B_{\rm c2}$(0) of the A15-type HEAs decrease, while the ratio of $B_{\rm c2}$(0)/$T_{\rm c}$ for 0.2 $\leq$ $x$ $\leq$ 1.2 is larger than or comparable to that of Nb$_{3}$Sn. The decrease in $T_{\rm c}$ is concomitant with a decrease in both the density of states at the Fermi level [$N$($E_{\rm F}$)] and electron-phonon coupling strength, as expected from the Bardeen-Cooper-Schreiffer (BCS) theory. In addition, the valence electron account (VEC) dependence of $T_{\rm c}$ for these A15-type HEAs is compared with those for the binary A15 and other structurally different HEA superconductors. The implication of these results is discussed.

\section{Experimental}
The V-Nb-Mo-Al-Ga HEAs were prepared by arc melting high purity V (99.9\%), Nb (99.9\%), Mo (99.9\%), Al (99.9\%) powders and Ga shots (99.9\%) according to the stoichiometric ratio of V:Nb:Mo:Al:Ga = (1.5-$x$/2):(1.5-$x$/2):$x$:0.5:0.5 with $x$ = 0.2, 0.4, 0.6, 0.8, 1.0, 1.2 and 1.4. In order to minimize the volatilization of Al and Ga during the arc melting process, the mixture was prereacted at 1000 $^{\circ}$C for one week, and then melted in an arc furnace under high-purity argon atmosphere. The melts were turned over and remelted several times to ensure homogeneity, followed by rapid cooling on a water-chilled copper plate. For annealing experiments, the as-cast ingots were thoroughly grounded and pressed into pellets, which were loaded in alumina crucibles in an argon-filled glove box. The crucibles were then placed in Ta tubes, sealed in evacuated quartz tubes and heated at 1600 $^{\circ}$C under argon atmosphere in a muffle furnace for 12 h. After furnace cooling to 900 $^{\circ}$C, the quartz tubes were quenched into the cold water.

The phase purity of as-cast and annealed HEAs was checked by power x-ray diffraction using a Bruker D8 Advance x-ray diffractometer with Cu K$\alpha$ radiation at room temperature. The structural refinements were performed using the JANA2006 program \cite{27JANA}. The chemical composition was measured with an energy-dispersive x-ray spectrometer (Model Octane Plus) affiliated to a Zeiss field emission scanning electron microscope. The spectra were collected on different locations of each sample for averaging. The electron diffraction was taken in a JEM-2100F transmission electron microscope operated at an accelerating voltage of 200 kV.
The electrical resistivity and specific heat were measured by the standard four-probe and the relaxation methods, respectively. Resisvity and specific heat measurements down to 1.8 K and up to 9 T were carried out in a Quantum Design PPMS-9 Dynacool. The dc magnetization measurements down to 1.8 K were performed in a commercial SQUID magnetometer (MPMS3).

\section{Results and Discussion}
\begin{figure}
\includegraphics*[width=9cm]{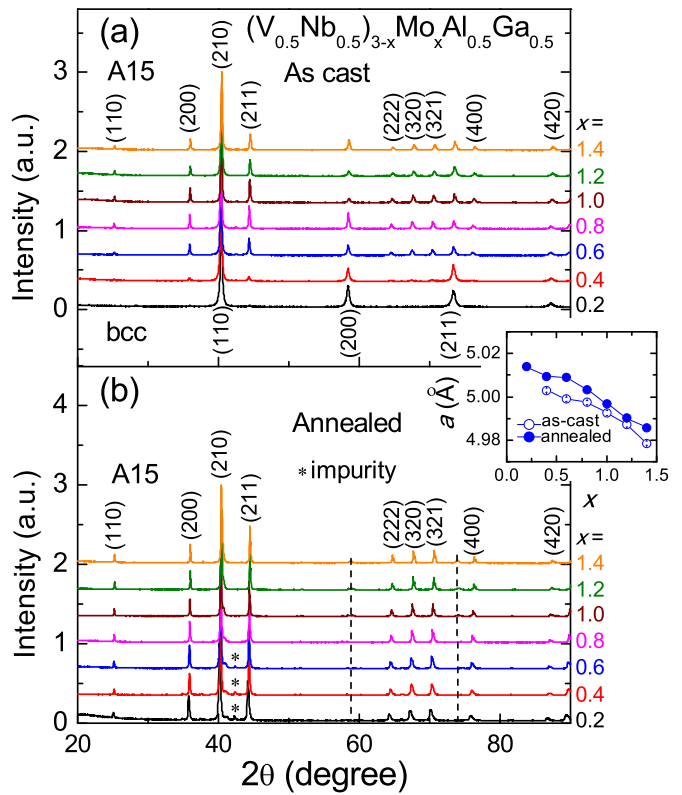}
\caption{(Color online)
(a), (b) Powder x-ray diffraction patterns at room temperature for the as-cast and annealed (V$_{0.5}$Nb$_{0.5}$)$_{3-x}$Mo$_{x}$Al$_{0.5}$Ga$_{0.5}$ HEAs. The major diffraction peaks for the as cast HEAs with $x$ $\leq$ 0.4 are indexed to the bcc structure. While for $x$ $\geq$ 0.6, the major peaks are indexed to the A15 structure. For annealed HEAs, the major peaks are indexed to the A15 structure, and the small impurity peaks for $x$ $\leq$ 0.6 are marked by the asterisks. The inset between (a) and (b) shows the lattice parameter of the A15 phase for the as-cast and annealed HEAs plotted as a function of the Mo content $x$.
}
\label{fig1}
\end{figure}

\begin{figure*}
\includegraphics*[width=15cm]{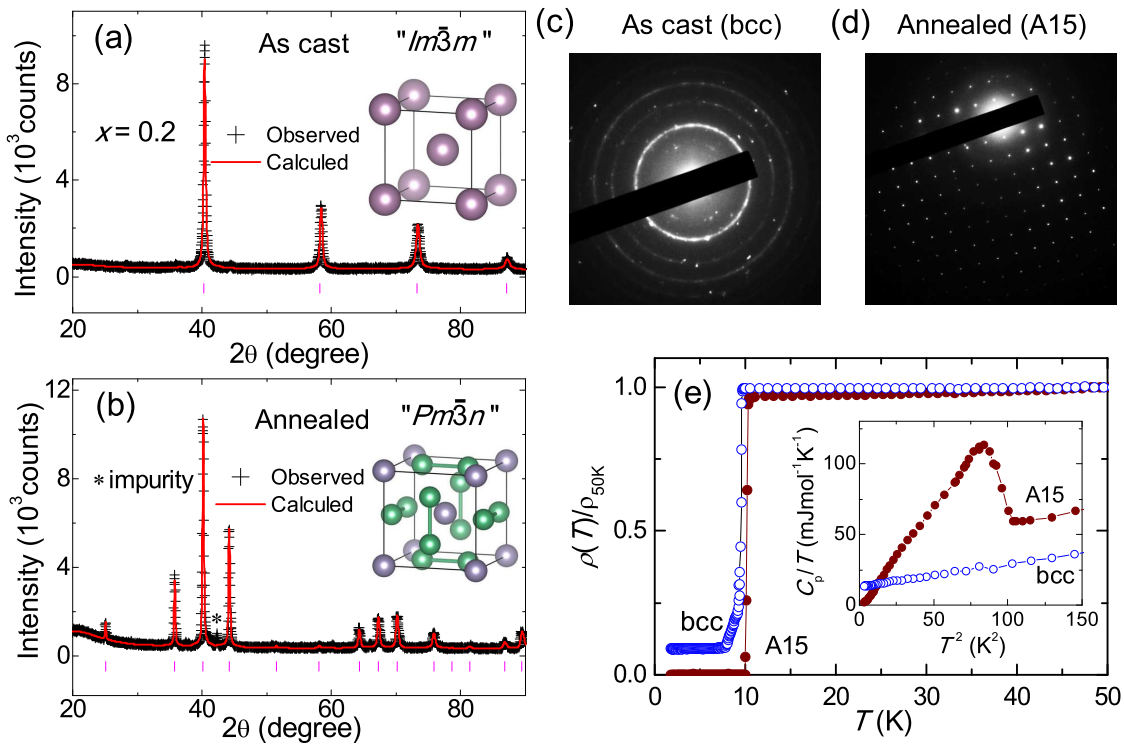}
\caption{(Color online)
(a),(b) The structural refinement profiles for the as-cast and annealed (V$_{0.5}$Nb$_{0.5}$)$_{3-x}$Mo$_{x}$Al$_{0.5}$Ga$_{0.5}$ HEAs with $x$ = 0.2. The small impurity peak in (b) is marked by the arrow. (c),(d) Electron diffraction patterns for the bcc and A15 polymorphs, respectively. (e) Temperature dependence of resistivity for the two different polymorphs. The inset shows the corresponding low-temperature specific heat data plotted as $C_{\rm p}$/$T$ versus $T^{2}$.
}
\label{fig2}
\end{figure*}

\begin{table*}
\caption{Refined crystallographic data of the as-cast and annealed (V$_{0.5}$Nb$_{0.5}$)$_{3-x}$Mo$_{x}$Al$_{0.5}$Ga$_{0.5}$ HEA with $x$ = 0.2.}
\renewcommand\arraystretch{1.3}
\begin{tabular}{p{3.5cm}<{\centering}p{0.5cm}<{\centering}p{1.6cm}<{\centering}p{0.5cm}<{\centering}p{4cm}<{\centering}p{0.5cm}<{\centering}p{1.6cm}<{\centering}p{0.5cm}<{\centering}p{4cm}<{\centering}p{2cm}<{\centering}}
\\
\hline 
&&As-cast &&&& Annealed &&\\
\hline 
Structural type &&bcc &&&& A15 &&\\
Space group && \textit{I}\textit{m}$\bar{3}$\textit{m} &&&& \textit{P}\textit{m}$\bar{3}$\textit{n} &&\\
Lattice parameter &&3.163 {\AA} &&&& 5.014 {\AA} &&\\
$R_{\rm wp}$ factor &&6.8\% &&&& 9.3\% &&\\
$R_{\rm p}$ factor  &&5.3\% &&&& 6.6\% &&\\
\hline 
   Atoms &  $x$  & $y$ & $z$ & Occ.&  $x$  & $y$ & $z$ & Occ. \\
V1/Nb1/Mo1/Al1/Ga1								 & 0	& 0 & 0 & 0.35/0.35/0.05/0.125/0.125& 0& 0& 0& 0.35/0.35/0.05/0.125/0.125 	\\
V2/Nb2/Mo2/Al2/Ga2								 &  &  &  & & 0.25& 0& 0.5& 0.35/0.35/0.05/0.125/0.125		\\
\hline
\end{tabular}
\label{Table3}
\end{table*}

\begin{figure*}
\includegraphics*[width=17cm]{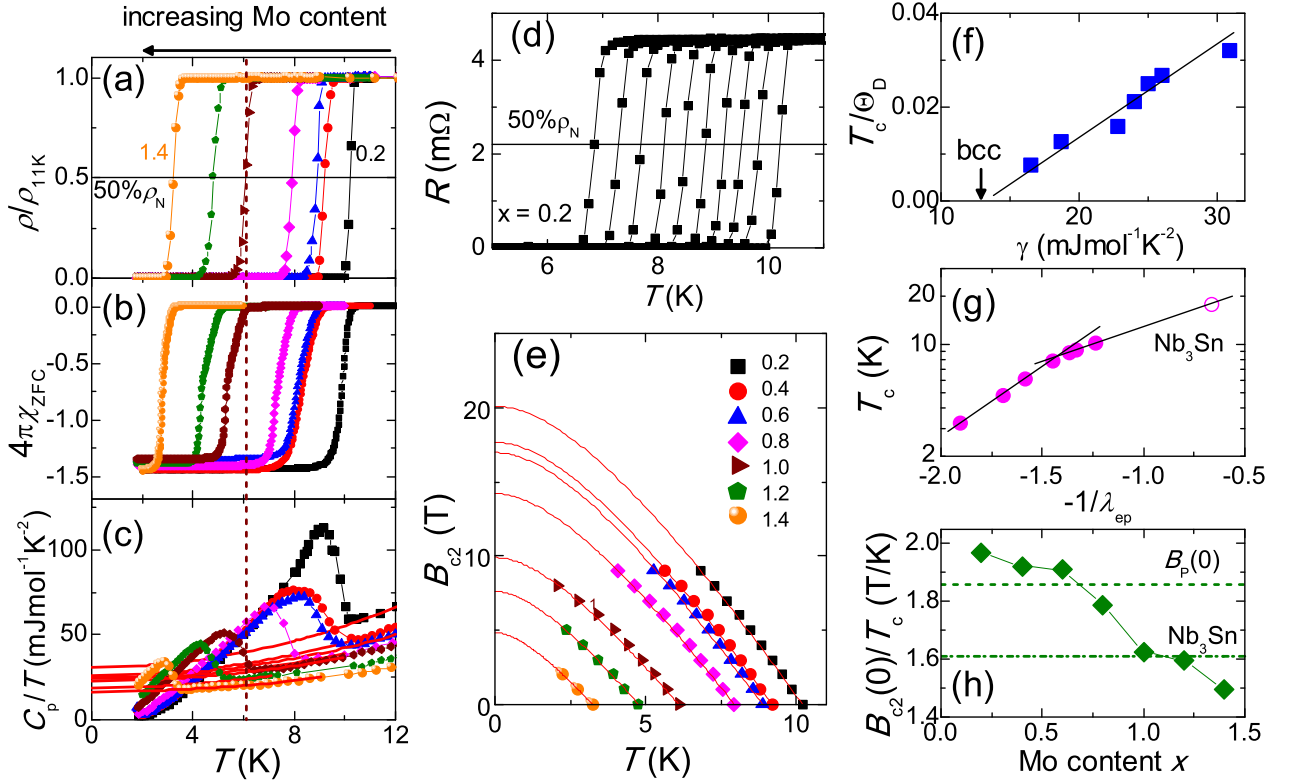}
\caption{(Color online)
(a)-(c) Temperature dependence of resistivity, magnetic susceptibility and specific heat below 12 K for the series of A15-type (V$_{0.5}$Nb$_{0.5}$)$_{3-x}$Mo$_{x}$Al$_{0.5}$Ga$_{0.5}$ HEAs. The left-handed arrow indicates the direction of increasing $x$, and the vertical dashed line is a guide to the eyes. In panel (a), the horizontal line denotes the midpoint of the resistive transition. In panel (c), the red solid lines are the fits by the Debye model to the normal-state data (see text for details). (d)  Temperature dependence of resistivity for $x$ = 0.2 under magnetic fields up to 9 T in increments of 1 T. The dashed line marks the midpoint of the resistive transition. (e) Temperature dependence of the upper critical fields determined for the series of HEAs. The solid lines are fits from the WHH model to the data. (f)  $T_{\rm c}$/$\Theta_{\rm D}$ plotted as a function of $\gamma$. The solid line is a guide to the eyes, and the arrow marks the $\gamma$ value for the bcc polymorph with $x$ = 0.2. (g) Logarithm of $T_{\rm c}$ plotted a function of $-$1/$\lambda_{\rm ep}$ for the HEAs, together with the data of Nb$_{3}$Sn. The solid lines are a guide to the eyes. (h) Mo content $x$ dependence of $B_{\rm c2}$(0)/$T_{\rm c}$. The dashed and dashed-dotted lines denote the values for the Pauli limiting field $B_{\rm P}$(0)/$T_{\rm c}$ = 1.86 T/K and Nb$_{3}$Sn, respectively.
}
\label{fig2}
\end{figure*}
The powder x-ray diffraction (XRD) patterns at room temperature for the series of as-cast and annealed (V$_{0.5}$Nb$_{0.5}$)$_{3-x}$Mo$_{x}$Al$_{0.5}$Ga$_{0.5}$ HEAs are shown in Figure 1(a) and (b), respectively. As can be seen from Fig. 1(a), for the as-cast HEA with $x$ = 0.2, only several diffraction peaks are observed and can be well indexed to the bcc structure with a lattice constant of 3.163 {\AA}. At higher $x$ values, however, a number of additional peaks start to appear at 2$\theta$ angles corresponding to the A15-type structure, whose strongest peak overlaps with that of the bcc one. The peak intensity of the A15 structure tends to grow with increasing $x$, yet the coexistence of two phases persists up to $x$ = 1.4, the highest Mo content investigated. Upon annealing at 1600 $^{\circ}$C, the peaks from the bcc structure disappears but those from the A15 structure remain. This reveals that the thermal treatment results in a bcc-to-A15 structural transformation. For $x$ $\leq$ 0.6, a small impurity peak is observed at 2$\theta$ $\approx$ 42$^{\circ}$, probably coming from the Nb$_{2}$Al-type sigma phase. By using a least-squares method, the lattice constant of the A15 structure for both the as-cast and annealed HEAs is determined and plotted as a function of the Mo content $x$ in the inset of Fig. 1. One can see that, for $x$ = 0.2, the $a$-axis is $\sim$5 {\AA}, and the corresponding unit-cell volume ($\sim$125 {\AA}$^{3}$) is nearly four times that of the bcc polymorph ($\sim$31.6 {\AA}$^{3}$). At higher $x$ values, the $a$-axis is 0.1-0.2\% larger in annealed HEAs than in as-cast HEAs. Nevertheless, in both cases, the $a$-axis decreases monotonically with increasing $x$, which means that the unit-cell volume shrinks as more Mo atoms are incorporated. On the other hand, energy dispersive x-ray analyses show that the chemical compositions of annealed (V$_{0.5}$Nb$_{0.5}$)$_{3-x}$Mo$_{x}$Al$_{0.5}$Ga$_{0.5}$ HEAs are essentially in agreement with the nominal ones, and more importantly, no composition change is observed before and after the annealing process for $x$ = 0.2 (see Table S1, Supporting Information).

\begin{table*}
\caption{Normal-state and superconducting parameters of A15-type (V$_{0.5}$Nb$_{0.5}$)$_{3-x}$Mo$_{x}$Al$_{0.5}$Ga$_{0.5}$ HEAs.}
\renewcommand\arraystretch{1.3}
\begin{tabular}{p{3.5cm}<{\centering}p{1.5cm}<{\centering}p{1.5cm}<{\centering}p{1.5cm}<{\centering}p{1.5cm}<{\centering}p{1.5cm}<{\centering}p{1.5cm}<{\centering}p{1.5cm}<{\centering}}
\\
\hline 
   & $x$ = 0.2   &  $x$ = 0.4  & $x$ = 0.6 & $x$ = 0.8 & $x$ = 1.0 & $x$ = 1.2 & $x$ = 1.4 \\

\hline 
$T_{\rm c}$ (K)							& 	  10.2	 & 9.2	& 8.9 & 7.9 & 6.1 & 4.8 & 3.2 \\
$\gamma$ (mJ mol$^{-1}$ K$^{-2}$)	&       30.9 & 26.0 & 25 & 24 & 22.8 & 18.7 & 16.5 \\
$\Delta$$C_{\rm p}$/$\gamma$$T_{\rm c}$							& 	  2.01 	 & 1.61 & 1.60 &1.54 & 1.31 & 1.33 & 1.33	\\
$\Theta_{\rm D}$ (K)				&      317   & 			342 & 355 & 374 & 383 & 377 & 424 	 \\
$\lambda_{\rm ph}$	&      0.81 &   0.75  & 0.73 & 0.69 & 0.63 & 0.59 & 0.52 \\
$B_{\rm c2}(0)$ (T)				&      20.1   & 			17.7& 17.0 & 14.2 & 9.9 & 7.6& 4.8	 \\
$\xi_{\rm GL}$ (nm)				&      4.0   & 			4.3 & 4.4 & 4.8 & 5.8 &6.6 & 8.3	 \\
\hline
\hline 
\end{tabular}
\label{Table3}
\end{table*}
From the above results, one can see that the (V$_{0.5}$Nb$_{0.5}$)$_{3-x}$Mo$_{x}$Al$_{0.5}$Ga$_{0.5}$ HEAs have two polymorphs at ambient pressure: one with the bcc structure and the other with the A15 structure. To gain more insight, we focus on the HEA with $x$ = 0.2, where the two polymorphs are isolated from each other. Figure 2(a) and (b) show the structural refinement profiles for the ac-cast and annealed HEAs at this composition based on the \textit{I}\textit{m}$\bar{3}$\textit{m} and \textit{P}\textit{m}$\bar{3}$\textit{n} space groups, respectively, and the refined results are listed in Table 1. In both cases, there is a good agreement between the observed and calculated XRD patterns, as indicated by the low $R_{\rm wp}$ and $R_{\rm p}$ values. In the bcc structure, there is only one crystallographic site for the five different type of atoms, whose occupancies are set by the respective atomic fraction. For the A15 structure, there are two distinct sites (0.25, 0, 0.5) and (0, 0, 0), while the atomic occupancies at each site remains unchanged. As a consequence, the one-dimensional chains consist of both transition metal and main group element atoms, in contrast to the binary A15 superconductors \cite{28A15structure}. Figure 2(c) and (d) show the electron diffraction patterns for the bcc and A15 polymorphs, respectively. In the former case, the pattern consists of a series of concentric rings, which match well the crystal planes of the bcc structure and is indicative of a small grain size.  By contrast, for the A15 one, well defined spots from (100) planes can be observed, which also provides evidence for the growth in grain size during the annealing process. Overall, these results are very similar to those observed in binary $A_{3}$$B$-type compounds with the bcc and A15 polymorphs \cite{25Nb3Al-1}. Nevertheless, it is noted that, for example, annealing Nb$_{3}$Al at 800 $^{\circ}$C for several hours is sufficient to transform the bcc polymorph to the A15 one completely. This heat-treatment temperature is only half that employed for the present HEAs. Actually, we have also performed annealing experiments at 1000 $^{\circ}$C, but found that the transformation is still not complete after more than one week. Hence it appears that, compared with binary compounds, the HEAs indeed have higher thermal stability. Furthermore, since the A15 polymorph can be regarded as a low-temperature phase and remains stable up to 1600 $^{\circ}$C, the polymorphic transition from the A15 to bcc structure should occur at a temperature much higher than 1600 $^{\circ}$C.

Recently, the effect of Al alloying on the bcc-type (TaNb)$_{0.67}$(ZrHfTi)$_{0.33}$ HEA superconductor has been investigated by von Rohr \emph{et al.} \cite{HEAAl}.
The results show that the [(TaNb)$_{0.67}$(ZrHfTi)$_{0.33}$]$_{1-x}$Al$_{x}$ remains in a bcc structure up to $x$ = 0.3, but changes to a sigma phase structure with increasing $x$ to 0.4, suggesting that the system is in the vicinity a structural instability.
In this respect, a polymorphic transition may occur for some $x$ value upon proper thermal treatment.
It is worth noting that, even for the Al-free Ta-Nb-Zr-Hf-Ti HEAs, the bcc structure can be unstable against annealing, though no other polymorph has been observed \cite{TaNbZrHfTi}.
This suggests that the occurrence of structure polymorphism in HEAs is closely related to their constituent elements.

Figure 2(e) shows the temperature dependence of normalized resistivity below 50 K for the bcc and A15 polymorphs of the (V$_{0.5}$Nb$_{0.5}$)$_{3-x}$Mo$_{x}$Al$_{0.5}$Ga$_{0.5}$ HEA with $x$ = 0.2. A weak temperature dependence is observed in both cases, which is typical for HEAs \cite{16bccHEASC,20HEABc2}. On cooling below 10.3 K, the resistivity of the A15 polymoprh drops sharply to zero, evidencing a superconducting transition. While a similar resistivity drop is seen at a slightly lower temperature for the bcc polymoprh, no zero resistance is achieved down to 1.8 K. On the other hand, a large specific heat ($C_{\rm p}$) jump is seen for the A15 polymoprh [see Fig. 2(e)], confirming the bulk nature of its superconductivity. Nevertheless, there is no anomaly in the $C_{\rm p}$ data of the bcc counterpart. This suggests that the bulk bcc phase does not superconduct above 1.8 K and the observed resistive transition originates from a tiny amount of superconducting impurities. In addition, the $C_{\rm p}$  data indicates that the electronic specific heat coefficient ($\gamma$) for the bcc polymoprh is much smaller than that of the A15 one, which will be discussed further below. In the remaining of the paper, we restrict our attention to the superconducting properties of the A15 polymorph in annealed (V$_{0.5}$Nb$_{0.5}$)$_{3-x}$Mo$_{x}$Al$_{0.5}$Ga$_{0.5}$ HEAs.

Figure 3(a)-(c) show the temperature dependence of resisitivity ($\rho$), magnetic susceptibility ($\chi$) and $C_{\rm p}$ of these HEAs below 12 K, respectively. For each $x$ value, the superconducting transition is characterized by a sharp drop in $\rho$, a strong diamagnetic response and a clear $C_{\rm p}$ jump. Here $T_{\rm c}$, determined as the midpoint of the resistive transition, is found to decrease monotonically from 10.2 K to 3.2 K with increasing $x$ to $x$ = 1.4. The onset temperature of the diamagnetic signal in the zero-field cooling $\chi_{\rm ZFC}$ agrees well with the $T_{\rm c}$ determined from the $\rho$ measurements. Moreover, the $\chi_{\rm ZFC}$ data at 1.8 K for all HEAs corresponds to a shielding fraction of more than 130 \% without correcting the demagnetization factor. This, together with the $C_{\rm p}$ jump, clearly demonstrates bulk superconductivity in these HEAs. In the normal state, the $C_{\rm p}$ data are well fitted by the Debye model $C_{\rm p}$/$T$ = $\gamma$$T$ + $\beta$$T^{2}$, where $\beta$ is the phononic specific heat coefficient. Once $\beta$ is known, the Debye temperature $\Theta_{\rm D}$ can be calculated by the formula $\Theta_{\rm D}$ = (12$\pi$$N$$R$/5$\beta$)$^{1/3}$, where $N$ is the number of atoms per unit cell and $R$ = 8.314 Jmol$^{-1}$K$^{-1}$ is the molar gas constant. The analysis results are summarized in Table 2. For $x$ = 0.2, $\gamma$ has a value of 30.9 mJmol$^{-1}$K$^{-2}$ and the normalized specific heat jumps $\Delta$$C_{\rm p}$/$\gamma$$T_{\rm c}$ is estimated to be 2.01, which is significantly larger than the BCS value of 1.43 \cite{29BCS}. Nevertheless, the jump can be reasonably reproduced by a modified BCS model \cite{alphamodel} with $\Delta_{0}$/$T_{\rm c}$ = 2.07, where $\Delta_{0}$ is the fully isotropic gap at 0 K (see Fig. S1, supporting information). This suggests that the superconducting state is still BCS-like. With increasing $x$, the two quantities decrease and reduce to 16.5 mJmol$^{-1}$K$^{-2}$ and 1.33 at $x$ = 1.4, respectively. On the other hand, with the knowledge of $T_{\rm c}$ and $\Theta_{\rm D}$, the electron-phonon coupling constant $\lambda_{\rm ep}$ can be calculated by the inverted McMillan formula, \cite{30Mcmillan}
\begin{equation}
\lambda_{\rm ph} = \frac{1.04 + \mu^{\ast} \rm ln(\Theta_{\rm D}/1.45\emph{T}_{\rm c})}{(1 - 0.62\mu^{\ast})\rm ln(\Theta_{\rm D}/1.45\emph{T}_{\rm c}) - 1.04},
\end{equation}
where $\mu^{\ast}$ is the Coulomb repulsion pseudopotential. Assuming that $\mu^{\ast}$ = 0.13, $\lambda_{\rm ep}$ values are also found to decreases from 0.81 to 0.52 with increasing $x$ from 0.2 to 1.4. Taken together, these results suggest that annealed (V$_{0.5}$Nb$_{0.5}$)$_{3-x}$Mo$_{x}$Al$_{0.5}$Ga$_{0.5}$ HEAs are moderately coupled superconductors for 0.2 $\leq$ $x$ $\leq$ 0.8 and weakly coupled superconductors for 1 $\leq$ $x$ $\leq$ 1.4.

To obtain the $B_{\rm c2}$(0) values for these HEAs, temperature dependent $\rho$($T$) was measured under various magnetic fields up to 9 T, an example of which for $x$ = 0.2 is shown in Fig. 3(d) (for other $x$ values see Figure S2, Supporting Information). With increasing magnetic field, the resistive transition is slightly broadened and shifts toward lower temperatures as expected. The $T_{\rm c}$ value at different fields is determined based on the same criterion as that under zero field, and the resulting $B_{\rm c2}$($T$) phase diagrams are displayed in Fig. 3(e). By using the Wathamer-Helfand-Hohenberg theory \cite{31WHH}, $B_{\rm c2}$ can be extrapolated to zero temperature, yielding $B_{\rm c2}$(0) = 20.1 T, 17.7 T, 16.5 T, 14.2 T, 9.9 T, 7.6 T, and 4.8 T for $x$ = 0.2, 0.4, 0.6, 0.8, 1.0, 1.2, and 1.4, respectively. From $B_{\rm c2}$(0), the Ginzburg-Landau (GL) coherence length $\xi_{\rm GL}$ can be calculated as $\xi_{\rm GL}$ = $\sqrt{\Phi_{0}/2\pi B_{\rm c2}(0)}$ , where $\Phi_{0}$ = 2.07 $\times$ 10$^{-15}$ Wb is the flux quantum. This gives a relatively short $\xi_{\rm GL}$ varying between 4.0 to 8.3 nm for these HEAs.

The examination of the correlation between the physical properties for these HEAs reveals the following salient features. First, as shown in Fig.3(f), the ratio of $T_{\rm c}$/$\Theta_{\rm D}$ increase as the increase of $\gamma$ and hence $N$($E_{\rm F}$),which is consistent with the BCS theory \cite{29BCS}. In this respect, the absence of superconductivity for the bcc polymorph of $x$ = 0.2 is likely due to the low $N$($E_{\rm F}$) since its $\gamma$ value (12.9 mJmol$^{-1}$K$^{-2}$) is only $\sim$40\% of that of the A15 one. Nevertheless, the possibility of a weakening of the electron-phonon coupling strength cannot be excluded. Second, when plotting the logarithm of $T_{\rm c}$ as a function of the inverted electron-phonon coupling constant $-$1/$\lambda_{\rm ep}$ [see Fig. 3(g)], two different slopes are discernible, which contrasts with the case of Heusler superconductors \cite{32Heusler}. Note that these two slopes correspond well to the weak (1 $\leq$ x $\leq$ 1.4) and moderate (0.2 $\leq$ x $\leq$ 0.8) electron-phonon coupling regimes found above, respectively. In addition, the data of the A15 superconductor Nb$_{3}$Sn falls on the extrapolation of the data in the moderate coupling regime. Nevertheless, the slope appears to be more steep for small $\lambda_{\rm ep}$, pointing to a more significant role of electron-phonon coupling on $T_{\rm c}$ in this regime. Third, as can be seen from Fig. 3(h), $B_{\rm c2}$(0)/$T_{\rm c}$ ratio of these HEAs is larger than or comparable to that of Nb$_{3}$Sn \cite{33Nb3SnBc2}, despite their much lower $T_{\rm c}$. Especially, the $B_{\rm c2}$(0) value for $x$ $\leq$ 0.6 slightly exceeds the Pauli paramagnetic limit, $B_{\rm P}$(0) = 1.86$T_{\rm c}$. While this $B$$_{\rm c2}$ behavior has been observed in bcc-type (TaNb)$_{0.16}$(HfZrTi)$_{0.84}$ HEA \cite{20HEABc2}, its $T_{\rm c}$ and $B_{\rm c2}$(0) values are only about half of those in the present case. In the dirty limit, we have $B_{\rm c2}$(0) = $-$0.693($d$$B_{\rm c2}$/$d$$T$)$_{T=T_{\rm c}}$$T_{\rm c}$ \cite{31WHH} and ($d$$B_{\rm c2}$/$d$$T$)$_{T=T_{\rm c}}$ $\propto$ $\gamma$$\rho_{\rm c}$ \cite{43dirtylimit}, where $\rho_{\rm c}$ is the normal-state resistivity value just above $T_{\rm c}$. Hence one can see that $B_{\rm c2}$(0)/$T_{\rm c}$ $\propto$ $\gamma$$\rho_{\rm c}$. Since the $\gamma$ values of A15-type HEAs are smaller than that of Nb$_{3}$Sn \cite{44gammaNb3Sn}, their larger $B_{\rm c2}$(0)/$T_{\rm c}$ is apparently attributed to an enhancement of $\rho_{\rm c}$ resulting from the strong disorder, similar to ball-milled Nb$_{3}$Sn \cite{35disorderBc22}. Also, it is worth noting that the $B_{\rm c2}$(0)/$T_{\rm c}$ ratio tends to increase with the increase of $T_{\rm c}$, and that a $B_{\rm c2}$(0) value of $\sim$20 T is attainable for a $T_{\rm c}$ of 10.2 K. Provided the $T_{\rm c}$ can be enhanced to $\sim$15 K, the $B_{\rm c2}$(0) of the A15-type HEAs might be above 30 T, which is higher than that of Nb$_{3}$Sn \cite{33Nb3SnBc2} and may have application in high field superconducting magnets.

\begin{figure}
\includegraphics*[width=8.7cm]{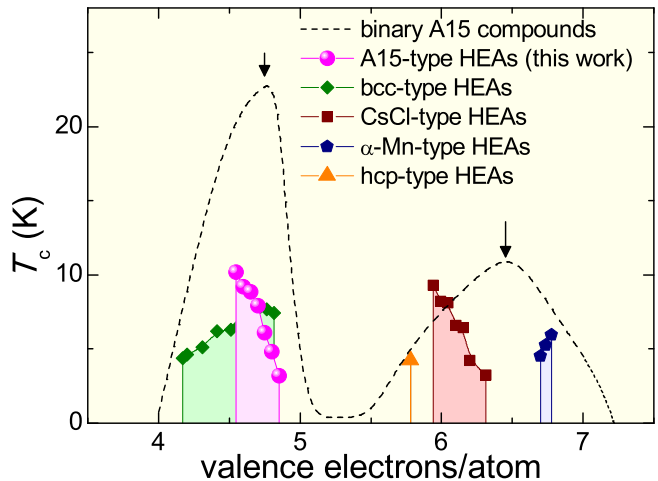}
\caption{(Color online)
Dependence of $T_{\rm c}$ on the average number of valence electrons per atom ratio for A15-type (V$_{0.5}$Nb$_{0.5}$)$_{3-x}$Mo$_{x}$Al$_{0.5}$Ga$_{0.5}$ HEAs. The data for binary A15 \cite{36A15SC} and other structurally different HEA \cite{16bccHEASC,17alphaMnHEASC,18CsClHEASC,19hexagonalHEASC} superconductors are also included for comparison. The solid lines are a guide to the eyes, and the two $T_{\rm c}$ maxima for binary A15 compounds are marked by the arrows.
}
\label{fig3}
\end{figure}
To provide hints for the $T_{\rm c}$ enhancement, we plot the $T_{\rm c}$ against the average number of valence electrons per atom ratio ($e$/$a$) for A15-type (V$_{0.5}$Nb$_{0.5}$)$_{3-x}$Mo$_{x}$Al$_{0.5}$Ga$_{0.5}$ HEAs, together with the data of all known binary A15 and other prototype HEAs superconductors for comparison \cite{36A15SC}. As expected from the Matthias rule \cite{37Matthiasrule}, $T_{\rm c}$ of the binary A15 superconductors exhibits two maxima around $e$/$a$ = 4.75 and 6.5, respectively. For the A15 HEAs, the $e$/$a$ values fall between 4.55 and 4.85, suggesting that they still obey the the Matthias rule. Nevertheless, their $T_{\rm c}$ shows a monotonic decrease with increasing $e$/$a$. This contrast is not so surprising since the strong disorder inherent to A15 HEAs will smear out $N$($E_{\rm F}$) and wipe out its structure that would be in the binary A15 superconductors \cite{38Labbemodel}. Hence the band filling dependence of $T_{\rm c}$ is naturally expected to be different between these two families of materials. In any case, this result suggests that the $T_{\rm c}$ of A15 HEAs could be further enhanced by reducing $e$/$a$ to below 4.55. In this regard, replacing Mo with Ta or group IVB elements (Ti, Zr, Hf) will be of interest for future studies. On the other hand, it is worth noting that, among all HEA superconductors, only the A15- and bcc-type ones exhibit an overlap of the $e$/$a$ ranges, which is consistent with the observation of a polymorphic transition between these two structural types.
Nevertheless, the $T_{\rm c}$ dependence on $e$/$a$ is quite different for two cases.
Hence it appears that, in addition to valence electron count, crystal structure and constituent elements also play a nontrivial role in determining the $T_{\rm c}$ of HEA superconductors.

Finally, we briefly discuss the implication of our results.
First, these results clearly indicate that HEA superconductors are not limited to transition metal elements, which open more possibilities of combinations of elements to compose superconducting HEAs.
Second, the A15-type HEAs represent an alternative way of introducing strong disorder to study the effect of atomic ordering on the physical properties of A15 compounds. Previously, such disorder can be introduced by ball milling \cite{40ballmilling}, rapid quenching \cite{41rapidquenching} and high-energy particle radiation \cite{42radiation}, all of which have no influence on the chemical composition. In this regard, the chemical complexity in HEAs is expected to provide a fresh insight into the effect of compositional disorder. For example, a natural question to ask is how this disorder affects normal state resistivity and magnetoresistance, which requires high magnetic fields to suppress superconductivity completely.
Third, it is prudent to note that the (V$_{0.5}$Nb$_{0.5}$)$_{3-x}$Mo$_{x}$Al$_{0.5}$Ga$_{0.5}$ HEAs exhibit the same polymorphism as the binary combinations of their constituent elements, such as Nb$_{3}$Al, Mo$_{3}$Al and V$_{3}$Ga.
Further studies are called for to assess the generality of this observation in other polymorphic HEAs.

\section{Conclusion}
In summary, we have discovered both superconductivity and temperature-driven polymorphism in the (V$_{0.5}$Nb$_{0.5}$)$_{3-x}$Mo$_{x}$Al$_{0.5}$Ga$_{0.5}$ HEAs with $x$ in the range 0.2 and 1.4.
The results show that as-cast HEA has a single bcc structure for $x$ = 0.2 and a mixture of the bcc and A15 structures for higher $x$ values.
Upon annealing, a bcc-to-A15 polymorphic transition takes place and all HEAs are found to exhibit bulk superconductivity. In particular, for $x$ = 0.2, while the bcc polymorph
is not superconducting down to 1.8 K, the A15 one has a $T_{\rm c}$ of 10.2 K and an estimated $B_{\rm c2}$(0) of 20.1 T, both of which are the highest among HEA superconductors. With increasing Mo content $x$, $T_{\rm c}$ of these A15-type HEAs decrease monotonically, which is ascribed to the reduction in both electron phonon coupling strength and $N$($E_{\rm F}$). Furthermore, the $T_{\rm c}$ also decreases with increasing $e$/$a$ from 4.55 to 4.85. This is different from the binary A15 as well as other structurally different HEA superconductors, and suggests that $T_{\rm c}$ may be enhanced by reducing the number of valence electrons.
On the other hand, the $B_{\rm c2}$(0)/$T_{\rm c}$ ratio of the A15-type HEAs, which is larger than that of Nb$_{3}$Sn over a broad $x$ range, provides evidence for a disorder-induced enhancement of the upper critical field. Our results not only present a new type of HEA superconductors, but also provide the first example of polymorphism dependent superconductivity in HEAs, which help to understand the interplay between chemical disorder, crystal structure and superconducting properties in these materials.
\\

\section*{ACKNOWLEGEMENT}
The work at Zhejiang University is financially supported by the National Key Research Development Program of China (No.2017YFA0303002).

\end{document}